\documentclass[useAMS]{mn2e}
\usepackage{epsfig,amssymb,times,color} 
\makeatletter
\def\figsize{\ifSFB@referee0.5\hsize\else\hsize\fi}
\makeatother

\def\mic {\hbox{$\mu$m}}
\def\eq#1{\begin{equation} #1 \end{equation}}
\def\E#1{\hbox{$10^{#1}$}}
\def\sub#1{_{\rm #1}}
\def\about {\hbox{$\sim$}}
\def\ga {\hbox{$\gtrsim$}}
\def\la {\hbox{$\lesssim$}}
\def\cc {\hbox{cm$^{-3}$}}
\def\x {\hbox{$\times$}}

\def\Mo {\hbox{M$_\odot$}}
\def\Lo {\hbox{L$_\odot$}}

\def\deg {\hbox{$^\circ$}}
\def\L {\hbox{$L_\star$}}
\def\T {\hbox{$T_\star$}}
\def\Tc {\hbox{$T_{\rm c}$}}
\def\Rc {\hbox{$R_{\rm c}$}}
\def\Rout {\hbox{$R_{\rm out}$}}
\def\Rcone {\hbox{$R_{\rm cone}$}}
\def\ta {\hbox{$\tau^{\rm a}_{\rm V}$}}
\def\te {\hbox{$\tau^{\rm e}_{\rm V}$}}

\def\R {\hbox{$R_\star$}}

\def\kms {\hbox{km s$^{-1}$}}

\newcount\hour
\newcount\minute
\def\clock {\hour = \time \divide\hour by 60 \minute = \hour
\multiply\minute by -60 \advance\minute by \time \number\hour:\ifnum\minute <
10 {0\number\minute} \else\number\minute \fi}
\def\Draft{}

\title[\Draft]
{Bipolar outflow on the Asymptotic Giant Branch---the case of IRC+10011}

\author[Vinkovi\'c et al.]
{Dejan Vinkovi\'{c},$^{1,2}$ Thomas Bl\"{o}cker,$^3$ Karl-Heinz
 Hofmann,$^3$ Moshe Elitzur$^2$
 \newauthor
 and Gerd Weigelt$^3$\\
$^1$School of Natural Sciences, Institute for Advanced Study,
Princeton, NJ 08540, USA; dejan@ias.edu\\
$^2$Department of Physics \& Astronomy, University of Kentucky,
Lexington, KY 40506-0055, USA; moshe@uky.edu\\
$^3$Max-Planck-Institut f\"{u}r Radioastronomie, Auf dem H\"{u}gel 69, 53121
Bonn, Germany;
khh@mpifr-bonn.mpg.de, weigelt@mpifr-bonn.mpg.de \\
}

\date{\Draft}
\pagerange{\pageref{firstpage}--\pageref{lastpage}} \pubyear{2004}

\begin{document}
\label{firstpage}

\maketitle

\begin{abstract}
Near-IR imaging of the AGB star IRC+10011 (= CIT3) reveals the presence of a
bipolar structure within the central \about\ 0.1 arcsec of a spherical dusty
wind. We show that the image asymmetries originate from \about\ \E{-4} \Mo\ of
swept-up wind material in an elongated cocoon whose expansion is driven by
bipolar jets. We perform detailed 2D radiative transfer calculations with the
cocoon modeled as two cones extending to \about\ 1,100 AU within an opening
angle of \about\ 30\deg, imbedded in a wind with the standard $r^{-2}$ density
profile. The cocoon expansion started $\la$ 200 years ago, while the total
lifetime of the circumstellar shell is \about\ 5,500 years. Similar bipolar
expansion, at various stages of evolution, has been recently observed in a
number of other AGB stars, culminating in jet breakout from the confining
spherical wind. The bipolar outflow is triggered at a late stage in the
evolution of AGB winds, and IRC+10011 provides its earliest example thus far.
These new developments enable us to identify the first instance of symmetry
breaking in the evolution from AGB to planetary nebula.

\end{abstract}


\section{Introduction}

The transition from spherically symmetric Asymptotic Giant Branch (AGB) winds
to non-spherical Planetary Nebulae (PNe) represents one of the most intriguing
problems of stellar astrophysics. While most PNe show distinct deviations from
spherical symmetry, their progenitors, the AGB stars, are conspicuous for the
sphericity of their winds (see, e.g., review by Balick \& Frank 2002). There
have been suggestions, though, that deviations from sphericity may exist in
some AGB winds, and perhaps could be even prevalent (Plez \& Lambert\ 1994,
Kahane at al.\ 1997). Thanks to progress in high resolution imaging, evidence
of asymmetry has become more conclusive for several objects in recent years (V
Hya: Plez \& Lambert\ 1994, Sahai et al.\ 2003a; X Her: Kahane \& Jura 1996;
IRC+10216: Weigelt et al. 1998 \& 2002, Haniff \& Buscher 1998, Skinner et al.
1998, Osterbart et al. 2000; RV Boo: Bergman et al.\ 2000, Biller et al.\ 2003;
CIT6: Schmidt et al.\ 2002).

The star IRC+10011 (= IRAS 01037+1219, also known as CIT3 and WXPsc), an
oxygen-rich long-period variable with a mean infrared variability period of 660
days (Le Bertre 1993), is one of the most extreme infrared AGB objects. This
source served as the prototype for the first detailed models of AGB winds by
Goldreich \& Scoville\ (1976) and of the OH maser emission from OH/IR stars by
Elitzur, Goldreich, \& Scoville\ (1976). The optically thick dusty shell
surrounding the star was formed by a large mass loss rate of \about \E{-5} \Mo
yr$^{-1}$. The shell expansion velocity of \about\ 20 \kms\ has been measured
in OH maser and CO lines. Various methods and measurements suggest a distance
to IRC+10011 in the range of 500 to 800 pc

For an archetype of spherically symmetric AGB winds, the recent discovery by
Hofmann et al.\ (2001; H01 hereafter) of distinct asymmetries in the IRC+10011
envelope came as a surprise. They obtained the first near infrared bispectrum
speckle-interferometry observations of IRC+10011 in the J-, H- and K'-band with
respective resolutions of 48 mas, 56 mas and 73 mas. While the H- and K'-band
images appear almost spherically symmetric, the J-band shows a clear asymmetry.
Two structures can be identified: a compact elliptical core and a fainter
fan-like structure. H01 also performed extensive one-dimensional radiative
transfer modelling to explain the overall spectral energy distribution (SED)
and angle-averaged visibility curves. Their model required a dust shell with
optical depth $\tau(0.55 \mu m)=30$ around a 2250 K star, with a dust
condensation temperature of \hbox{900 K}. This one-dimensional model
successfully captured the essence of the circumstellar dusty environment of
IRC+10011 but could not address the observed image asymmetry and its variation
with wavelength. In addition, the model had difficulty explaining the far-IR
flux, requiring an unusual transition from a $1/r^2$ density profile to the
flatter $1/r^{1.5}$ for $r$ larger than 20.5 dust condensation radii. Finally,
the model produced scattered near-IR flux in excess of observations.

We report here the results of 2D radiative transfer modelling of IRC+10011 that
successfully explain the observed asymmetries. After analyzing in \S2 general
observational implications we describe in \S3 our model for a bipolar outflow
in IRC+10011. In \S4 we present detailed comparison of the model results with
the data and resolution of the problems encountered by the 1D modelling. The
discussion in \S5 advances arguments for the role of bipolar jets in shaping
the circumstellar envelope of IRC+10011 and other AGB stars. We conclude with a
summary in \S6.

\section{Observational Implications}

The near-IR images, especially the J-band, place strong constraints on the dust
density distribution in the inner regions. Emission at the shortest wavelengths
comes from the hottest dust regions. For condensation temperature \about\ 1,000
K the peak emission is at \about\ 4\mic, declining rapidly toward shorter
wavelengths. At 1.24 \mic, the J-band is dominated by dust scattering. It is
easy to show that scattering by a $1/r^p$ dust density distribution produces a
$1/r^{p+1}$ brightness profile (Vinkovi\'c et al, 2003). The J-band image from
H01 is elongated and axially symmetric. The decline of brightness from its
central peak along this axis of symmetry is different in the opposite
directions. In one direction it declines as $1/r^3$, corresponding to the
$1/r^2$ density profile typical of stellar winds. But in the other direction
the brightness falls off only as $1/r^{1.5}$, corresponding to the flat,
unusual $1/r^{0.5}$ density profile.

The large scale structure is not as well constrained by imaging. However, all
observations are consistent with the following simple picture: An optically
thick spherical wind has the standard $1/r^2$ density profile. Since the
buildup of optical depth is concentrated in the innermost regions for this
density law, the near-IR imaging penetrates close to the dust condensation
region. The wind contains an imbedded bipolar structure of limited radial
extent and density profile $1/r^{0.5}$. The system is observed at an
inclination from the axis so that the wind obscures the receding part of the
bipolar structure, creating the observed asymmetry of the scattering image,
which traces directly the density distribution. The inclination angle must be
$\la$ 45\deg\ since a larger value starts to expose the receding part. But the
inclination cannot be too small because the approaching part would get in front
of the wind hot dust, leading to a strong 10 \mic\ absorption feature, contrary
to observations. Because of its shallow density profile, the column density of
the bipolar structure {\em increases} as $r^{0.5}$ away from the condensation
cavity, and the size of J-band image corresponds to the distance where the
scattering optical depth reaches unity. Regions further out do not show up
because of self-absorption. Dust emission is affected also by the temperature
distribution, and the central heating by the star tends to produce spherical
isotherms. Images taken at longer wavelengths, such as the K-band, can thus
appear more symmetric.

Some qualitative estimates of the gas density follow immediately. The wind
optical depth at the J-band must be $\ga$ 1. This optical depth is accumulated
close to the dust condensation radius, roughly 3\x\E{14} cm for a distance of
650 pc. Assuming a standard dust-to-gas mass ratio of 1:100, the gas density at
the condensation radius is \ga\ 3\x\E7\ \cc. For the bipolar structure, the
J-band optical depth is \about\ 1 across the size of the observed image, which
is \about\ 2\x\E{15} cm. This leads to a density estimate of \about\ 7\x\E6\
\cc\ at the condensation radius within the bipolar structure. These rough
estimates are within a factor 10 of the results of the detailed modelling
described below.

The density at the base of the outflow is about an order of magnitude lower in
the bipolar structure than in the wind region. An outflow can bore its way
through another denser one only if its velocity is higher so that it plows its
way thanks to its ram pressure. The propagation of such high-velocity bipolar
outflows has been studied extensively in many contexts, beginning with jets in
extragalactic radio sources (Scheuer 1974). The jet terminates in a shock,
resulting in an expanding, elongated cocoon similar to the observed bipolar
structure. With a $1/r^{0.5}$ density law, most of the bipolar structure mass
is concentrated at its outer edge with the largest $r$, consistent with the
structure of the expanding cocoon.

\begin{figure}
\begin{center}
\framebox[3\width]{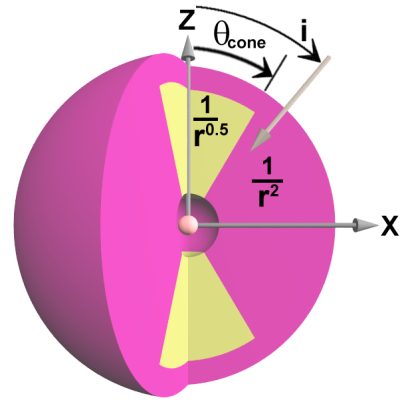}
\end{center}
\caption{\label{Model} Sketch of the 2D model for the circumstellar dusty shell
around IRC+10011. In a spherical wind with the standard $1/r^2$ density profile
are imbedded two polar cones with half-opening angle $\theta_{cone}$ and a
$1/r^{0.5}$ density profile. The system is viewed from angle $i$ to the axis.
}
\end{figure}

\section{2D Modeling of IRC+10011}

\begin{figure*}
\begin{minipage}{\textwidth}
\centering \leavevmode \epsfig{file=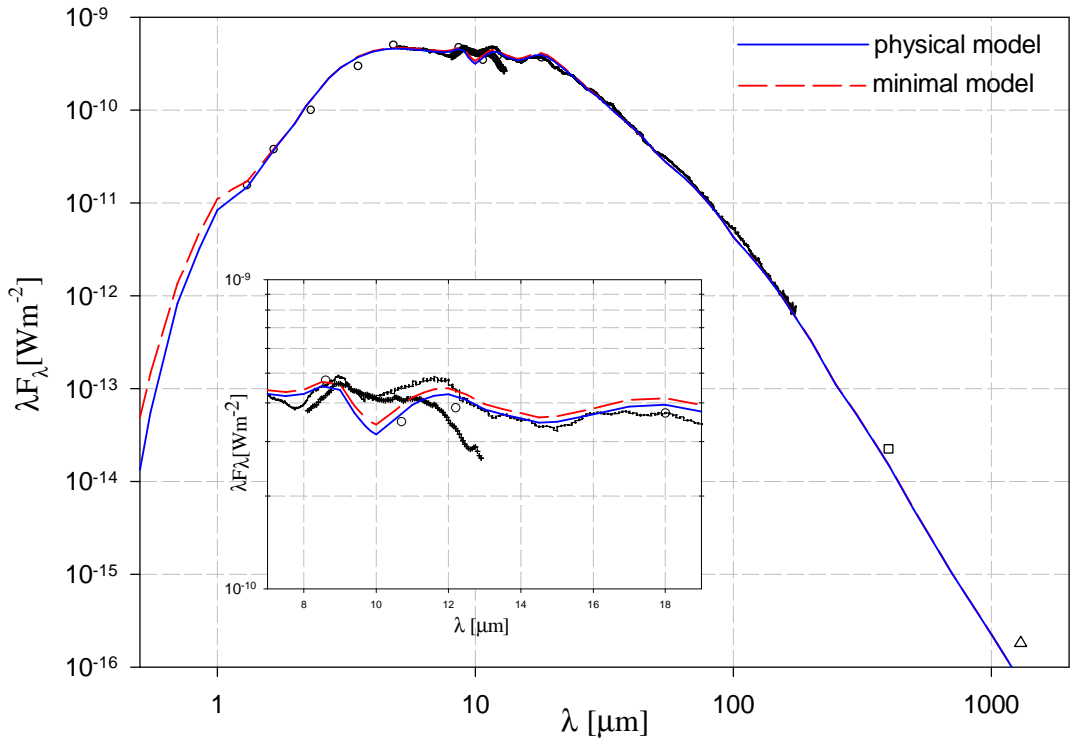,width=0.8\textwidth,clip}
\caption{\label{SEDfit} SED observations and modeling. Data (see H01) are
indicated with various symbols and lines. Thick, smooth lines show the SED
produced by two models: the dashed line corresponds to the minimal model
(\S\ref{sec:minimal}), the solid line to the final, physical model
(\S\ref{sec:physical}). The inset shows an expanded view of the 10\mic\ region.
All subsequent figures refer to the final physical model, whose properties are
summarized in Table 1.}
\end{minipage}
\end{figure*}
\begin{figure*}
\begin{minipage}{\textwidth}
\centering \leavevmode
\epsfig{figure=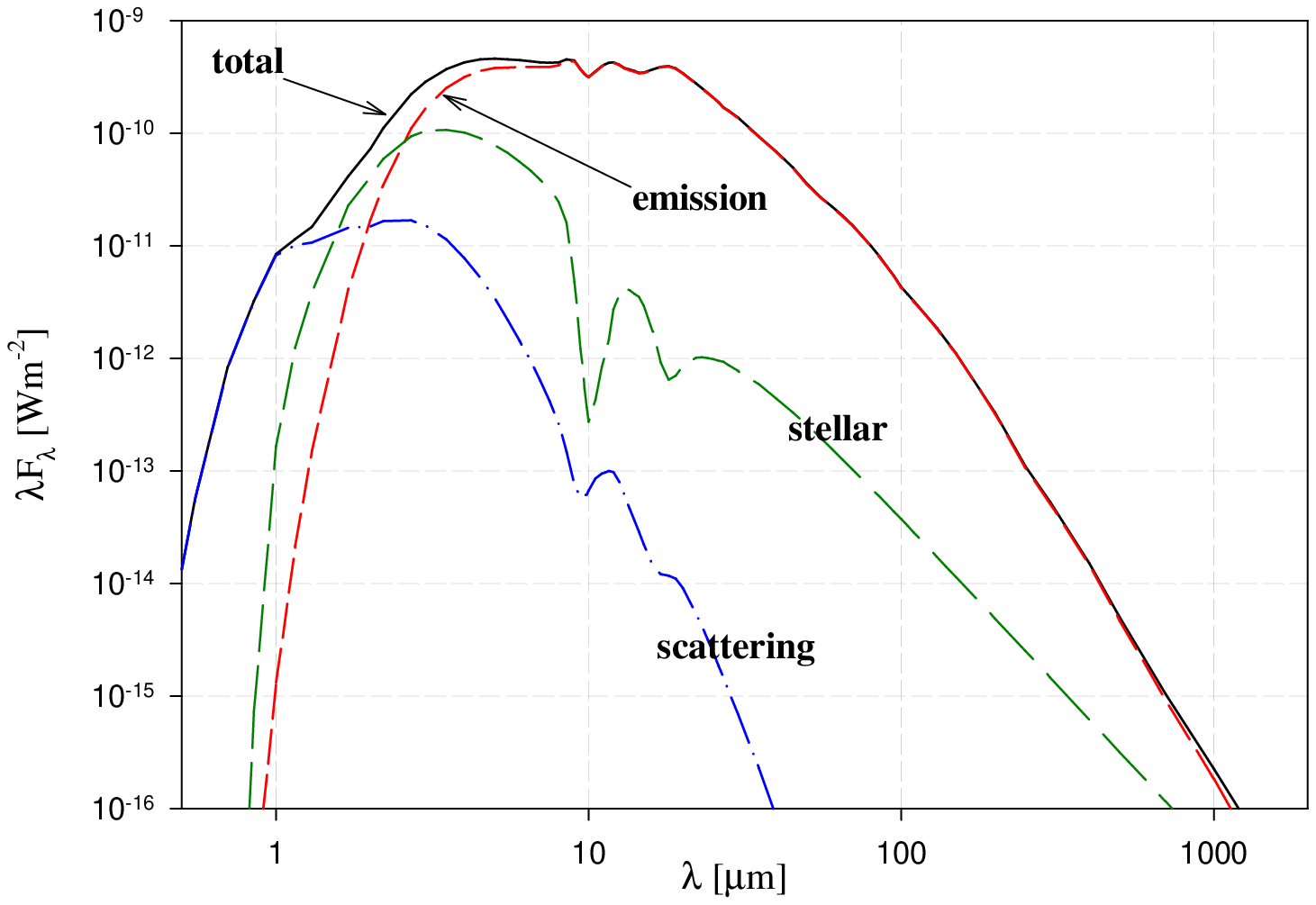,clip,width=0.48\textwidth}\hspace{0.5cm}
\epsfig{figure=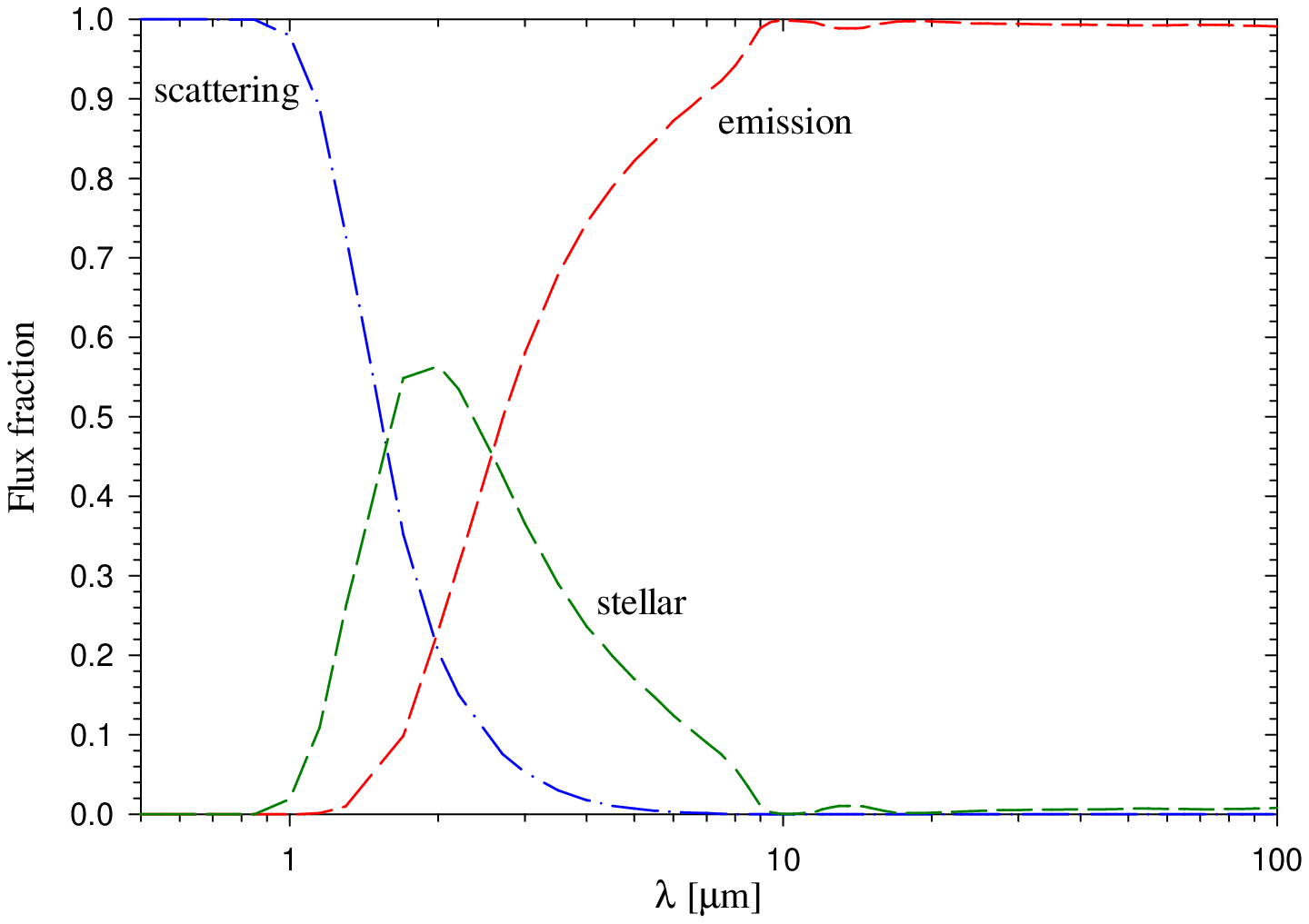,clip,width=0.48\textwidth}
\caption{\label{Flux_components} Left: The model SED and its
breakup to the stellar, dust scattering and emission components,
as indicated. Right: Wavelength variation of the relative
contribution of each component to total flux. Note the fast change
from scattering to emission dominance around 2\mic. This
transition is responsible for the observed wavelength variation of
the image asymmetry in the near-IR.}
\end{minipage}
\end{figure*}

\begin{figure}
\centering \epsfig{figure=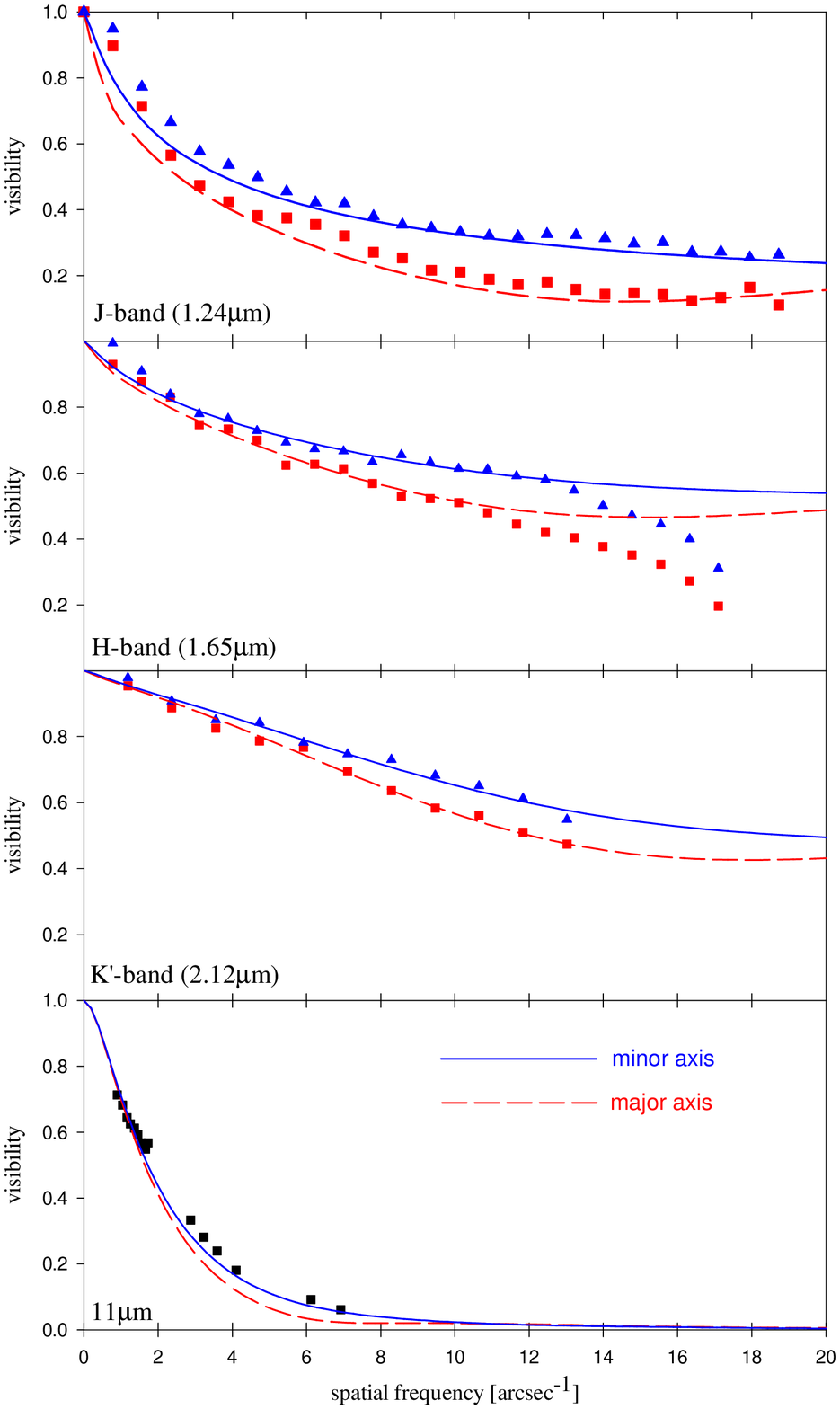,clip,width=\hsize}
\caption{\label{visibilities} Visibility functions. Lines are
model predictions, symbols are data points from H01 (near-IR) and
Lipman et al.\ 2000 (11 \mic). }
\end{figure}

Detailed 2D radiative transfer modeling was conducted with our newly developed
code LELUYA (see Appendix). Starting with the minimal configuration that
explains all available observations, we show on physical grounds that this
minimal model must be modified and propose a suitable modification.

\subsection{A Minimal Model}
\label{sec:minimal}

For the first working model we adopt the geometry shown in figure \ref{Model},
which requires the {\em minimal number of free parameters}. Each polar cone is
described by its half-opening angle $\theta\sub{cone}$ and radial extent
\Rcone. Apart from discontinuities across the cone boundaries, the density
depends only on $r$. It varies as $1/r^{0.5}$ inside the cones and $1/r^2$
outside, out to some final radius \Rout. To complete the description of the
geometry we need to specify its inner boundary, and it is important to note
that this cannot be done a-priori. Dust exists only where its temperature is
below the condensation temperature \Tc. Following H01 we select \Tc\ = 900 K.
The dust inner boundary, corresponding to the radial distance of dust
condensation, \Rc, is determined from \eq{\label{eq:Rc} T(\Rc(\theta)) = \Tc }
The equilibrium dust temperature, $T$, is set by balancing its emission with
the radiative heating. But the latter includes also the diffuse radiation,
which is not known beforehand when the dust is optically thick; it can only be
determined from the overall solution. Furthermore, because the spherical
symmetry is broken by the cones, the shape of the dust condensation surface can
be expected to deviate from spherical and is not known a-priori. Therefore
equation \ref{eq:Rc} completes the description of the geometry with an implicit
definition of the inner boundary $\Rc(\theta)$.

The radiative transfer problem for radiatively heated dust possesses general
scaling properties (Ivezi\'{c} \& Elitzur 1997). As a result, \Tc\ is the only
dimensional quantity that need be specified. All other properties can be
expressed in dimensionless terms. Luminosity is irrelevant, the only relevant
property of the stellar radiation is its spectral shape, which we take as
black-body at \T\ = 2,250 K. For individual dust grains, the only relevant
properties are the spectral shapes of the absorption and scattering
coefficients. For these we adopt spectral profiles corresponding to the
silicate grains of Ossenkopf, Henning \& Mathis (1992) with the standard size
distribution described by Mathis, Rumpl \& Nordsieck (1977; MRN).  Our
calculations employ isotropic scattering. These properties are the same
everywhere.

Density and distance scales do not enter individually, only indirectly through
overall optical depth. With two independent density regions, the problem
definition requires two independent optical depths. For this purpose we choose
\ta\ and \te, the overall optical depths at visual wavelengths along the axis
and the equator, respectively. Spatial dimensions can be scaled with an
arbitrary pre-defined distance, which we choose as the dust condensation radius
in the equatorial plane, \Rc(90\deg). Radial distance $r$ is thus replaced with
$\rho = r/\Rc(90\deg)$ so that, e.g., $\rho_{out} = \Rout/\Rc(90\deg)$.
Equation \ref{eq:Rc} becomes an equation for the scaled boundary of the
condensation cavity. The relation between angular displacement from the star
$\vartheta$ and the distance $\rho$ is \eq{\label{eq:theta} \vartheta =
{\vartheta_\star \over 2\rho_\star} \rho } where $\vartheta_\star$ is the
stellar angular size and $\rho_\star = \R/\Rc(90\deg)$ is the scaled stellar
radius. Physical dimensions can be set if one specifies a stellar luminosity
\L, which determines the condensation radius \Rc(90\deg).

\begin{figure*}
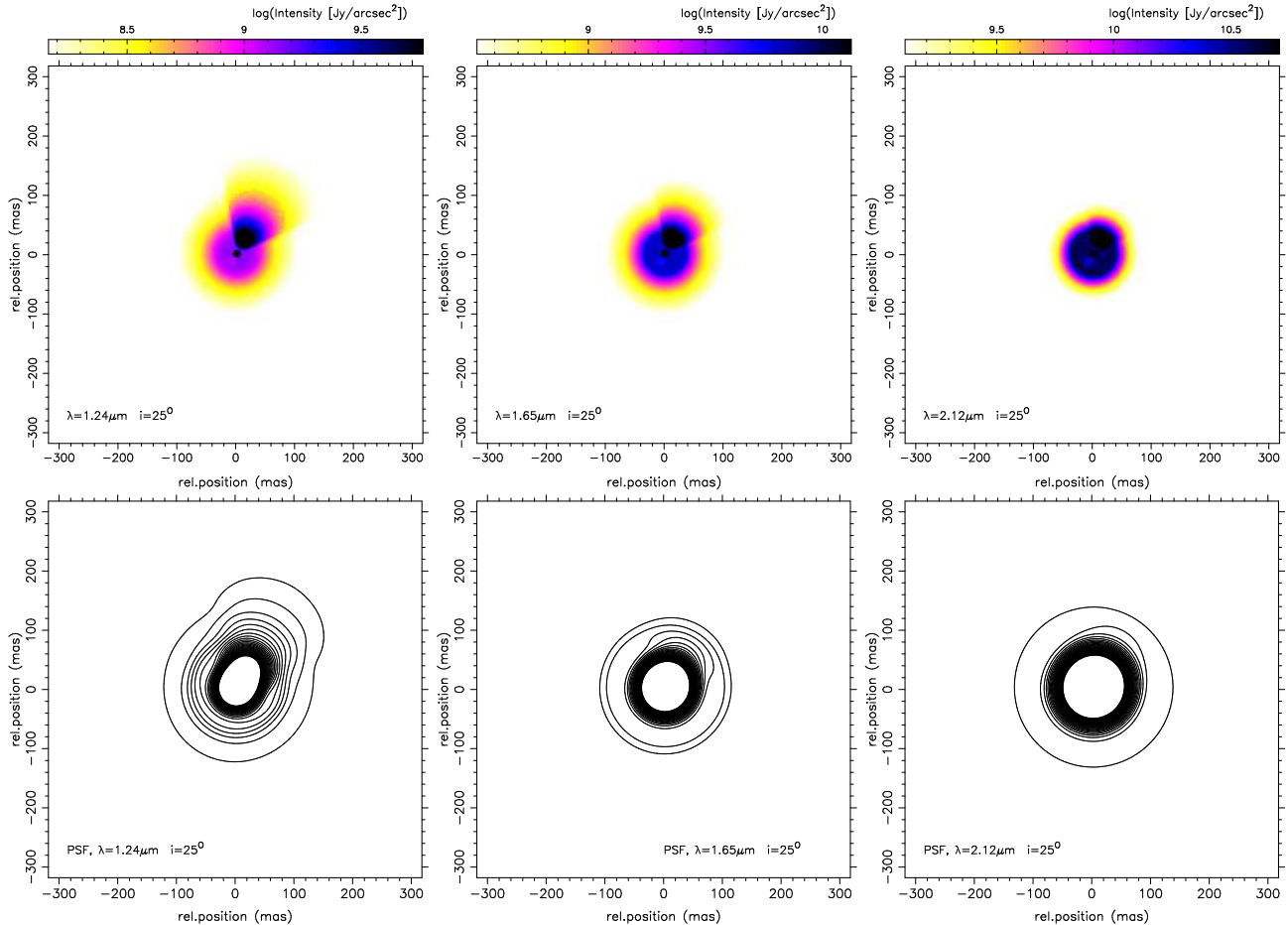

\begin{minipage}{\textwidth}
\centering \leavevmode
\epsfig{figure=CIT3_Jband_image.ps,clip,width=0.32\textwidth}
\epsfig{figure=CIT3_Hband_image.ps,clip,width=0.32\textwidth}
\epsfig{figure=CIT3_Kband_image.ps,clip,width=0.32\textwidth}
\epsfig{figure=CIT3_Jband_image-PSF.ps,clip,width=0.32\textwidth}
\epsfig{figure=CIT3_Hband_image-PSF.ps,clip,width=0.32\textwidth}
\epsfig{figure=CIT3_Kband_image-PSF.ps,clip,width=0.32\textwidth}
\caption{Theoretical J-band (1.24$\mu$m), H-band (1.65$\mu$m), and
K'-band (2.12$\mu$m) images of IRC+10011. Upper row: images for
perfect resolution, without PSF convolution. The dot at the center
of each image is the star. The nearby bright fan-shaped structure
is scattered light escaping through the cone. Lower row: Images
convolved with the instrumental PSF of H01. Contours are plotted
from 1.5\% to 29.5\% of the peak brightness in steps of 1\%. The
transition from scattered light dominance in the J-band to thermal
dust emission in the K'-band creates a sudden disappearance of the
image asymmetry.}\label{images}
\end{minipage}
\end{figure*}

To summarize, in all of our model calculations the following quantities were
held fixed: grain properties, \Tc\ = 900 K, \T\ = 2,250 K and the outer
boundary $\rho_{out} = 1000$. We varied \ta, \te, $\theta_{cone}$ and
$\rho_{cone}$. Once a model is computed, comparison with observations
introduces one more free parameter, the viewing angle $i$. A detailed
discussion of the data is available in H01. Figure \ref{SEDfit} shows the SED
for the best fit model which has \te\ = 40, \ta\ = 20, $\rho_{cone}=700$, and
$\theta_{cone}=15\deg$. The 10\mic\ region is difficult to fit in full detail.
Any further improvement would probably require more complicated geometry and/or
modified dust properties. The fit yields a bolometric flux of $F_{\rm bol} =
10^{-9}\,\rm W/m^2$, corresponding to $\vartheta_\star$ = 10.82~mas for the
stellar angular size, similar to the 10.9~mas derived in H01. The IR-flux
measurements determine the total amount of emitting dust. Assuming a standard
$n_d\sigma_d/n$ = \E{-21} cm$^{2}$, the overall mass of the IRC+10011
circumstellar shell is 0.13 \Mo.

\subsubsection{Shortcomings of the Minimal Model}
\label{subsection:shortcomings}

The minimal-model success in fitting the SED implies that it contains the
proper amount of dust. In addition to the SED, this model reproduces adequately
all the imaging observations at near-IR. However, these observations probe only
the innermost regions of the bipolar structure and do not provide adequate
constraints on its full extent.

We now show that the cones cannot extend all the way to $\rho\sub{cone}$ = 700,
as required from the fit to the SED by the minimal model. If the cones were
that large, the ratio of mass contained in them and in the wind region would be
$M\sub{cone}/M\sub{wind}$ = 1.7, that is, most of the circumstellar mass would
be in the cones. But such large mass cannot be swept-up wind material because
the fractional volume occupied by the cones is only 0.034. And building up this
mass with enhanced outflow through the polar regions can also be ruled out, as
follows: Mass conservation along stream lines yields $v_1 t = R_1\int\eta\rho^2
d\rho$, where $\eta(\rho) = n(\rho)/n_1$ is the dimensionless density profile,
$v_1$ and $n_1$ are the velocity and density at the streamline base $R_1$, and
$t$ is the duration of the outflow. Applying this relation to streamlines in
the cone ($\eta = \rho^{-1/2}$) and wind ($\eta = \rho^{-2}$) regions yields
\eq{\label{eq:vt}
 {v_1t\over R_1}\big\vert\sub{cone} =
   \frac25\rho_{\rm cone}^{5/2}, \qquad {v_1t\over R_1}\big\vert\sub{wind} =
   \rho\sub{out}
}
With $\rho\sub{cone}$ = 700, the product $v_1t/R_1$ is 5.2\x\E6\ in the cone
regions while in the wind it is only 1000. Since the wind starts with a sonic
velocity $v\sub{1w}$ \about\ 1 \kms, the conical outflow would have to start
with velocity $v\sub{1c} \simeq 5.2\x\E3\, t\sub{w}/t\sub{c}$ \kms, where
$t\sub{w}$ and $t\sub{c}$ are the wind and cones lifetimes. This is impossible
since the bipolar structure would extend much further than the wind even for
$t\sub{c} = t\sub{w}$; taking a physical $t\sub{c} \ll t\sub{w}$ only makes
things worse. This argument can be easily extended to show that, irrespective
of the magnitude of $\rho\sub{cone}$, the mass in the cones could not be
deposited purely by recent enhancement of polar mass loss rates. A substantial
fraction, perhaps even all, of this mass must be swept-up wind material.

\subsection{A Physical Model}
\label{sec:physical}

While our minimal model provides successful fits for all observations,
$\rho\sub{cone}$ cannot be as large as this model requires. Unfortunately,
observations do not yet meaningfully constrain $\rho\sub{cone}$ because at
$\rho$ \ga\ 10 the near-IR brightness drops below current detection
capabilities. A physical model for the origin of the bipolar structure
(\S\ref{sec:jet}) suggests that the cones extend to $\rho\sub{cone}$ = 47
(equation \ref{rho=47}). Adopting this value, the minimal model must be
modified to account for the far-IR flux produced by the large mass removed from
the rest of the cones. This mass can be placed elsewhere as long as its
temperature distribution corresponds to far-IR wavelengths (equation A7 in
Vinkovi\'{c} et al. 2003). With the total far-IR flux as the only observational
constrain, the only firm limit on this cold dust component is that it starts at
$\rho\ \ga$ 100 so that its temperature is $\la$ 100 K; in all other respects,
the geometry is arbitrary.

To account for the far-IR excess, the H01 wind model, which did not incorporate
the bipolar component, placed additional cold dust in a spherical component
with $\rho^{-1.5}$ density profile at $\rho\ \ga\ 20$. Here we propose a
simpler alternative: a detached spherical shell of increased dust density due
to an earlier phase of higher mass loss rate. The radial wind dust density
profile jumps by factor 3 at $\rho = 100$ so that
 \eq{
   \eta_{\rm wind} = \rho^{-2}\times
   \left\{\begin{array}{cl}
     1  & \hbox{for}\ \rho < 100 \\
     3  & \hbox{for}\ 100\le \rho \le 1000\\
   \end{array}\right.
 }
With the cold dust displaced from the cones to the wind, the equatorial optical
depth increases from \te\ = 40 in the minimal model to 40.72. Along the axis,
\ta\ drops from 20 to 5.1, of which 4.3 comes from the cones. As is evident
from figure \ref{SEDfit}, the SED produced by this physical model is almost
identical to that of the minimal model. The contributions of different
components to the total flux are shown in the left panel of figure
\ref{Flux_components}, with the fractional contributions shown in the right
panel.

Table 1 summarizes the input parameters, and various properties derived below,
of our model. The proposal of a two-shell model is motivated mostly by physical
plausibility since other than the SED, current observations do not place
meaningful constraints on the cold dust configuration. We have verified that a
disk geometry for the cold dust would also successfully reproduce the SED by
modeling with disk structures of $\rho=100$ inner radius and radial density
profiles $\rho^{-1/2}$ and $\rho^{-2}$. However, a disk geometry for the cold
dust component suffers from the same shortcomings as the extended cones of the
minimal model (\S\ref{subsection:shortcomings}).

The actual dust configuration is probably more complicated than our simple
description. Imaging observations at 8.55 \mic\ with spatial resolution of
$\rho$ \about\ 50 by Marengo et al.\ (1999) suggest an extension along an axis
almost perpendicular to the symmetry axis of the bipolar structure. This
possible additional asymmetry is not accounted by our physical model.

\begin{table}
\begin{center}
\begin{tabular}{lr}
\hline \hline
   stellar temperature             & 2250 K                \\
   stellar size                    & 10.8 mas (7 AU\dag)   \\
   luminosity\dag                  & 1.3$\cdot$\E4 \Lo     \\
   bolometric flux                 & \E{-9} W m$^{-2}$     \\
   dust condensation temperature   & 900 K                 \\
   condensation radius \Rc(90\deg) & 35.4 mas (23 AU\dag)  \\
   viewing angle $i$               & 25\deg                \\
   \hline
   Wind --- Inner shell:     &                           \\
   density profile           & $r^{-2}$                  \\
   inner boundary            & \Rc($\theta$)     (eq. 1) \\
   outer boundary            & 100\Rc(90\deg)            \\
   radial $\tau_{\rm V}$     & 39.6                      \\
   mass$^*$                  & 0.005 \Mo                 \\
   age\dag\dag               & 550 years                 \\
   mass loss rate$^*$\dag\dag& 9$\cdot$\E{-6} \Mo\ yr$^{-1}$\\
   \hline
   Wind --- Outer shell:     &                           \\
   density profile           & $r^{-2}$                  \\
   inner boundary            & 100\Rc(90\deg)            \\
   outer boundary            & 1000\Rc(90\deg)           \\
   radial $\tau_{\rm V}$     & 1.1                       \\
   mass$^*$                  & 0.13 \Mo                  \\
   age\dag\dag               & 5500 years                \\
   mass loss rate$^*$\dag\dag& 2$\cdot$\E{-5} \Mo\ yr$^{-1}$\\
    \hline
   Cone Properties:          &                           \\
   opening angle 2$\theta_{cone}$ & 30\deg                    \\
   density profile           & $r^{-0.5}$                \\
   outer boundary            & 47\Rc(90\deg)             \\
   axial $\tau_{\rm V}$      & 4.3                       \\
   mass$^*$                  & \E{-4} \Mo                \\
   age$^*$\dag               & $\la$ 200 years           \\
   \hline\hline
\end{tabular}
\caption{Model parameters and derived properties. Quantities
marked by \dag\ assume a distance of 650 pc, subject to an
uncertainty of $\pm$150 pc. Quantities marked by \dag\dag\
additionally assume a wind velocity of 20\,\kms. Quantities marked
by $^*$ assume $n_d\sigma_d/n$ = \E{-21} cm$^{2}$. Uncertainties
and the acceptable range of the various properties are discussed
in the text. }
\smallskip

\end{center}
\end{table}

\section{Visibility Functions and Images} \label{sec:Imaging}

With the model parameters set from the SED, the surface brightness distribution
is fully determined, and the visibility functions are calculated from the
brightness. For comparison with observations, the visibility must be normalized
with the flux collected within the field of view $\Theta\sub{FV}$. If the image
is divided into $N\x N$ pixels then the spatial frequency is $q_i =
i/\Theta\sub{FV}$, where $i=1...N$. Adopting $N = 300$, sufficiently large to
resolve the highest measured spatial frequencies, and $\Theta\sub{FV}$ from the
instrumental system, the results shown in figure \ref{visibilities} contain no
additional free parameters. In contrast with the SED, the visibility displays a
strong sensitivity to the grain size. A change of only 0.05 \mic\ in the
maximum grain size $a\sub{max}$ has a significant effect on the visibility
curves. Our physical model has $a_{max}$ = 0.20 \mic, resulting in good fits
for both the SED (figure \ref{SEDfit}) and the four different visibility curves
(figure \ref{visibilities}).

The J-band visibility is the most difficult to model because it is dominated by
the scattered light and thus very sensitive to fine details of the density
distribution and grain size. Since the agreement between data and theory is
better for small scales (higher spatial frequency), the quality of the fit to
the J-band image can be expected to deteriorate with distance from the star.
The model does not explain the puzzling drop in the H-band visibility at $q
\gtrsim 14$ cycles per arcsec, corresponding to structure smaller than the
condensation cavity. Since a similar drop is not present in the J-band, it must
correspond to material that emits but does not scatter light significantly. Hot
gas might be a possible explanation.

\begin{figure}
\begin{center}
\framebox[3\width]{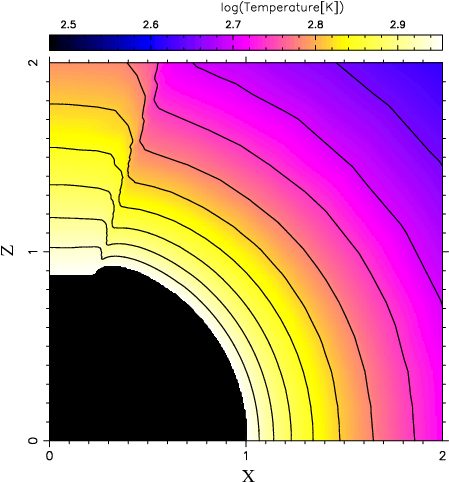}
\end{center}
\caption{\label{2Dtemperature} Temperature distribution around the condensation
cavity. The contours start at 850 K and decrease at 50 K intervals. The dust
condensation temperature is 900 K.}
\end{figure}

Our model images and their convolution with the instrumental PSF of H01 are
shown in figure \ref{images}. The comparison between the model and observed
images is satisfactory, indicating that the overall geometry is properly
captured by our simple model. The ``halo'' around the star in J-band model
image is brighter than observed. Possible explanations are dust accumulation
close to the equatorial region as well as asymmetric dust scattering. The
overall image asymmetry is much more prominent in the J-band, where dust
scattering dominates the radiative transfer (see figure \ref{Flux_components}).
As the wavelength shifts toward dominance of dust thermal emission, the image
becomes more symmetric. The reason is that scattered light traces directly the
density distribution while the dust emission is affected also by the
temperature distribution. Figure \ref{2Dtemperature} shows the dust temperature
distribution around the condensation cavity. Because of the central heating,
the temperature decreases with radial distance and tends to create circularly
symmetric isotherms. The asymmetric diffuse radiation distorts the isotherms,
but the deviations from circularity are small, especially at the high dust
temperatures traced by the K-band image. As a result, the image becomes more
symmetric, especially after convolution with the PSF as shown in the lower
panel of figure \ref{images}.

\begin{figure}
\begin{center}
\epsfig{figure=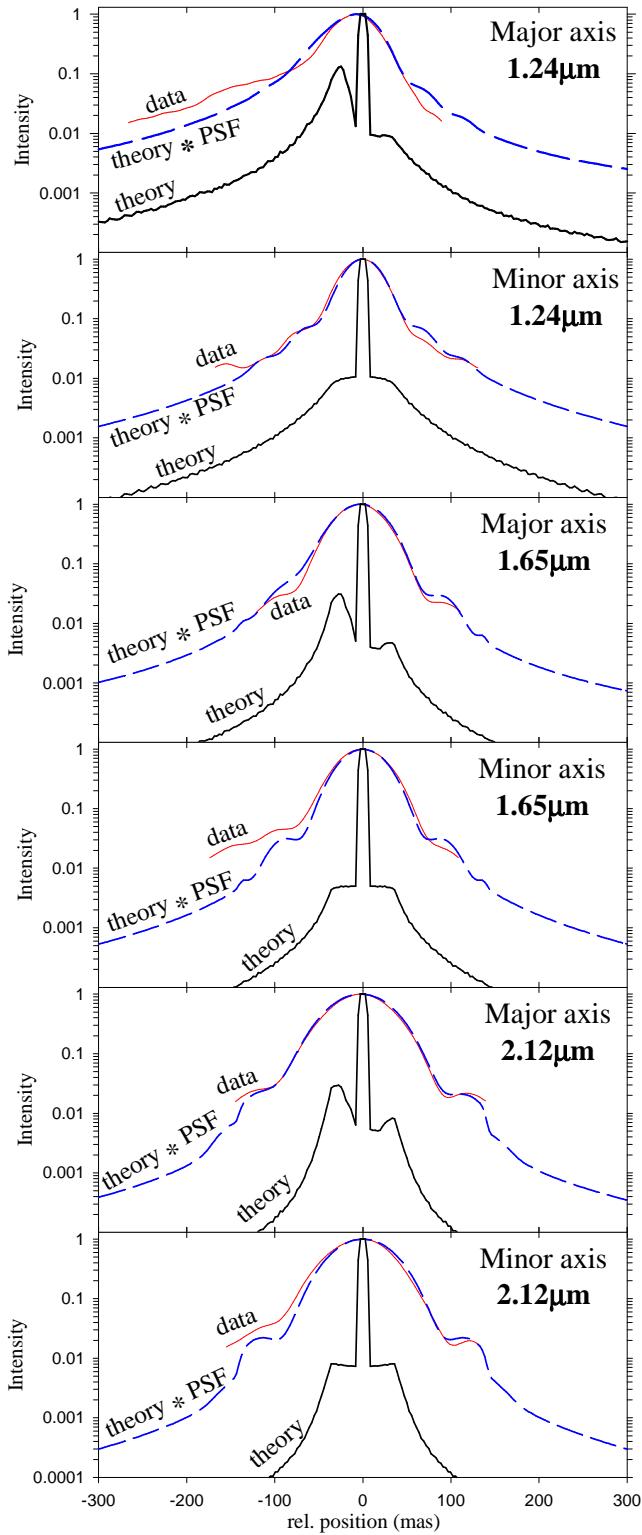,clip,width=\hsize}
\end{center}
\caption{\label{Image_profile}  Brightness profiles along the major and minor
axes in J (top two panels), H (middle two panels) and K (bottom two panels).
Thick lines show the model predictions with and without PSF convolution. The
thin lines show the profiles from the H01 data above the noise level (within
1.5\% of the peak brightness). The strong central peak in the theoretical
profile is the star, while the secondary peak visible on the major axis is
light scattered from the polar cone.}
\end{figure}

As evident from figure \ref{images}, the PSF convolution smears out the star
and the nearby fan-shaped structure into one broad elongated peak whose center
is shifted from the stellar position. This shift is more clearly noticeable in
the brightness profiles, shown in figure \ref{Image_profile}. The shift is 8.3
mas along the major axis in the J-band and 2.8 mas for the H- and K'-bands. The
images provide tight constraints on the inclination angle. Neither $i =
20^{\circ}$ nor $i = 30^{\circ}$ produce acceptable fits, so that
$i=25^{\circ}\pm 3^{\circ}$.

\section{Discussion}

Thanks to the scaling properties of dust radiative transfer,
neither luminosity, distance or density absolute scales were
specified. The distance to the source of 650$\pm$150 pc fixes
those scales, so that the luminosity is 1.3\x\E4\,\Lo\ and the
dust condensation radius is $R_c(90\deg) = 23\pm 5$ AU. The wind
inner shell extends to 2,300 AU, thus containing all the masers,
including the OH 1612 MHz (Elitzur, Goldreich \& Scoville 1976),
and its mass is 5\x\E{-3} \Mo. With a wind velocity of 20 \kms,
the duration of this phase is 550 years and the corresponding mass
loss rate is 9\x\E{-6} \Mo\ yr$^{-1}$. The wind outer shell
extends to 23,000 AU and its mass is 0.13 \Mo. Assuming the same
wind velocity, the duration of this phase is 5,500 years and the
corresponding mass loss rate is 2\x\E{-5} \Mo\ yr$^{-1}$.

Modelling uncertainties allow a few times larger size of the inner
shell, resulting in its larger mass and duration. The outer shell
has much smaller overall uncertainties since its total mass is
constrained by the far-IR flux. We can still derive general
conclusions that:
 \begin{enumerate}
 \item the duration of the inner shell is $\lesssim$1,000 years, while
 the outer shell is 5 to 10 times older;
 \item the overall dust opacity comes mostly from the inner shell;
 \item the mass is contained almost exclusively in the outer
 shell, with about 10 to 30 times more mass than the inner shell;
 \item the mass loss rate is of the order of \E{-5} \Mo\
 yr$^{-1}$, with a few times larger rate in the other shell;
 \item the overall circumstellar mass of
0.13 \Mo\ indicates that IRC+10011 is close to the end of its AGB
evolution.
 \end{enumerate}

\subsection{Dust Properties}

Our models employ silicate grains from Ossenkopf et al. (1992) with the
standard MRN size distribution. We found that the upper limit on the grain
sizes had to be reduced to $a\sub{max}$ = 0.20 \mic\ from the standard 0.25
\mic. While this change made little difference in the SED analysis, it was
necessary for proper fits of the visibility curves. The most important effect
of $a\sub{max}$ is control of the crossover from scattering to emission
dominance, crucial for explanation of the observed change from elongated to
circular images between the J- and K-bands (see \S\ref{sec:Imaging}). Although
we cannot claim to have determined the precise magnitude of $a\sub{max}$, the
fact that it is smaller than the standard seems certain.

The dust properties in our model were the same everywhere to minimize the
number of free parameters. In a detailed study of the proto-planetary nebula
IRAS 16342-3814, Dijkstra et al.\ (2003) find that the maximum grain size
varies from \about\ 1.3 \mic\ in a torus around the star to \about\ 0.09 \mic\
in the bipolar lobes. If such a variation in dust properties can occur already
on the AGB, the $a\sub{max}$ we find would represent an average over the cones
and wind regions.

\subsection{Jet Model for the Bipolar Structure}
\label{sec:jet}

The near-IR brightness observations map a region of the cones that extends to
$\rho \sim 8$ and has optical depth $\tau_V \sim 1.4$, corresponding to a gas
density at the base of each cone of $n\sub{1c}$ = 1.3\x\E6\ cm$^{-3}$. In
contrast, the gas density at the base of the wind region (obtained from \te\ =
41) is $n\sub{1w}$ = 1.7\x\E8\ cm$^{-3}$. The large density disparity amplifies
our earlier conclusion that the bipolar cones are sustained by high-velocity
ram pressure.

The small density at the base of the cones shows that their material has been
evacuated and deposited at larger distances by a recent event. We propose the
following simple scenario for the bipolar structure: High-velocity low-density
jets were recently turned on at the polar regions. The jets cleared out polar
cavities but are trapped by the material pushed ahead by their ram pressure,
resulting in an expanding cocoon as described first by Scheuer (1974). Our
model cones are a description of the current density distribution of the
cocoon, a snapshot of an inherently dynamic structure. In this picture, the
mass in the cones is swept-up ambient wind material and the cone boundary is
then
\eq{\label{rho=47}
 \rho\sub{cone} = \left(\frac52{n\sub{1w}\over n\sub{1c}}\right)^{2/3} = 47.
}
The swept-up mass is only \about\ \E{-4} \Mo. The leading edge of the cocoon
moves at velocity $v\sub{c} = \beta v\sub{w}$, where $v\sub{w}$ is the local
wind velocity and $\beta > 1$. From pressure balance during jet confinement,
\eq{
n\sub{c}(v\sub{c} - v\sub{w})^2 = n\sub{w}v_{\rm t}^2
}
where $n\sub{c}$ and $n\sub{w}$ are densities across the cocoon leading edge
and $v\sub{t}$ is the local speed of sound in the wind. This condition requires
that the density of the cones be smaller than the ambient density into which
they are expanding, i.e., $n\sub{c} < n\sub{w}$, restricting the cone radial
extension to $\rho \le$ 26 which is slightly smaller than the derived
$\rho\sub{cone}$. We attribute this discrepancy to the approximate nature of
our model, in which the complex structure of the cocoon--wind boundary is
replaced with the sharp-cutoff of the simple power-law density distribution of
the cones. Taking $n\sub{c} \simeq n\sub{w}$ at the cone boundary, pressure
balance implies $v\sub{c} \simeq v\sub{w}$, consistent with a recent start of
the jet confinement. Assuming that the cocoon radial boundary moves according
to $\rho\sub{cone} \propto t^{\alpha}$, with $\alpha$ \ga\ 1 to ensure
acceleration, its velocity is $v\sub{c} = \alpha\rho\sub{cone}R_1/t$. This
yields an estimate for the jet lifetime
\eq{
   t\sub{jet} = {\alpha\over\beta}
   {R_1\over v\sub{w}}\rho_{\rm cone}\ \la\ 200\ \rm years
}
for $\alpha/\beta\ \la$ 1.

Because of the steep decline of the wind density, the expansion
accelerates rapidly as the cocoon boundary reaches lower density
regions. Eventually it will break out of the wind, exposing the
underlying jets. Indeed, a striking example of such a
configuration comes from the recent observations, including proper
motion measurements, of water masers in W43A by Imai et al.\
(2002). The observations reveal tightly collimated velocities of
\about\ 150 \kms\ at distances up to \about\ 0.3 pc at the two
ends of an axis through the star. These masers are created by the
impact of the jets on clumps in the surrounding medium. In
addition to these far-away high-velocity masers, the source
displays the usual configuration typical of OH/IR stars -- OH and
water masers in shells expanding with velocities \about\ 9 \kms\
with radii of \about\ 500 AU. Therefore this source displays both
the spherical AGB wind and the jets that broke through it. A
similar, and probably more evolved, example is IRAS16342-3814.
Similar to W43A, this is a ``water-fountain'' jet PPN source but
its bipolar cavities are also seen with HST (Sahai et al. 1999).

\subsection{Asymmetry Evolution in AGB Stars}
\label{sec:sequence}

\begin{figure}
\begin{center}
\epsfig{figure=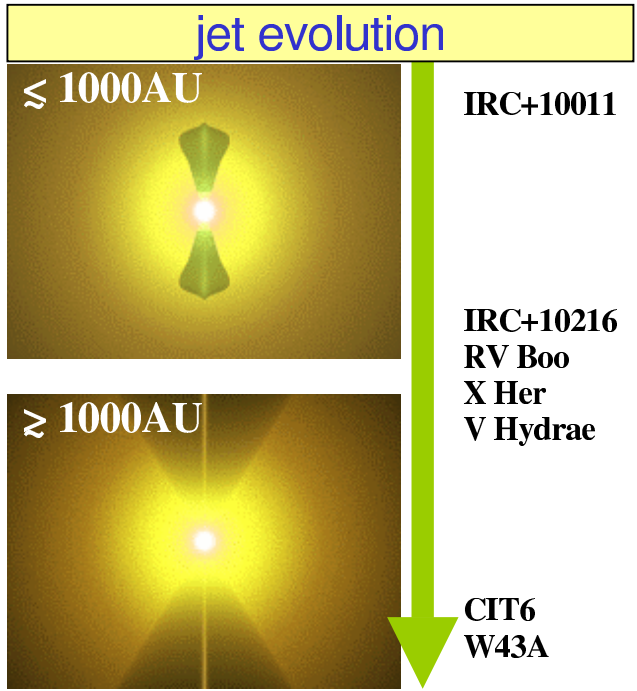,clip,width=\hsize}
\end{center}
\caption{\label{fig:Jet} Schematic representation of our proposed bipolar jet
evolution during the late stages of AGB winds. The top panel shows an early
phase when the jets are still confined by the spherical wind. IRC+10011 is the
earliest manifestation of this structure. The bottom panel illustrates evolved
jets that have broken out of the spherical confinement. CIT6 and W43A show
clear evidence of such structure. Several other sources give indications of
intermediate stages and are listed according to approximate ordering suggested
by current observations. The source RV Boo (Bergman et al. 2000, Biller et al.
2003) gives an indication of bipolarity, which is less certain than in the
other objects. }
\end{figure}

IRC+10011 and W43A can be considered, respectively, the youngest and most
evolved examples of sources displaying the evolution of bipolar jets working
their way through AGB winds. The proposal that jets, operating at the late AGB
or early post-AGB phase, are the primary mechanisms for shaping PNe has been
made already by Sahai \& Trauger (1998) and since supported by numerous
observations. The prototype C-rich star IRC+10216 shows circular shape on the
20 arcsec scale both in V-band (de Laverney 2003) and molecular line images
(e.g., Dayal \& Bieging 1995). But high-resolution IR imaging at the
0.1 arcsec scale reveal elongated structure similar to that in IRC+10011
(Osterbart et al.\ 2000, Weigelt et al.\ 2002). Unlike IRC+10011, though, where
only the J-band image gives clear indication of asymmetry, in IRC+10216 it is
evident even in the K-band. This strongly suggests that IRC+10216 represents a
more advanced stage than IRC+10011 of the evolution of a jet-driven cocoon
confined by the ambient spherical wind.

The C-rich star V Hya provides an example that is further along in evolution.
Recent CO observations by Sahai et al.\ (2003a) show that the bulk of the
emission comes from an elongated structure centered on the star. In addition,
an emission blob is approaching at a projected line-of-sight velocity of 250
\kms\ along an axis perpendicular to this elongation. This is the expected
morphology of a bipolar outflow breaking from the confinement of the
high-density region of the AGB wind if the receding blob is obscured by the
central torus. A similar structure has been found in the O-rich star X Her.
Partially resolved CO observations by Kahane \& Jura 1996 reveal a spherical
component expanding with only 2.5 \kms\ and two symmetrically displaced 10
\kms\ components, likely to be the red and blue shifted cones of a weakly
collimated bipolar flow. The bipolar lobes are \about\ 1.5 times bigger than
the spherical component. Finally, the C-rich star CIT6 presents an even more
evolved system. A bipolar asymmetry dominates the image in molecular line
mapping by Lindqvist et al.\ (2000), Keck imaging by Monnier et al.\ (2000) and
HST-NICMOS imaging by Schmidt et al.\ (2002).

Figure \ref{fig:Jet} shows a schematic sketch of our proposed evolutionary
sequence. The evolutionary stage of each indicated object is our rough estimate
based on current observations. This figure is only meant as an illustration of
the time-line suggested by our proposed scenario. The placing of different
objects is based on different kinds of data. For example, for X Her we only
have single-dish mm-wave observations with their attendant large beams, whereas
IRC+10216 and V Hya were observed with the much higher resolution of HST. Also,
whereas there is direct kinematic evidence for the high-velocity jets in V Hya
and W43A, the same is lacking for the other objects. We can also expect that in
objects with different jet properties (e.g., mass flux, speed, opening angle)
and different AGB mass-loss rates in the inner region, the jets will follow different
time histories of when they ``break out'' and the opening angle of the bipolar
cone which they dig in the AGB envelope will be different. Also, if
jets are episodic than they can change their direction (for which there is observational
evidence). The actual picture can be expected to be quite more complex than the
simple sketch in Fig \ref{fig:Jet}. Nevertheless, we expect the displayed
sequence to provide useful guidance for future studies.

\section{Summary and Conclusions}

We find that the circumstellar shell of IRC+10011 contains \about\ 0.13 \Mo,
extends to a radial distance of \about\ 23,000 AU (\about\ 35\arcsec) from the
star and is \about\ 5,500 years old. Most of the mass (\about\ 96\%) is
contained in the outer shell from \about\ 2,300 AU (\about\ 3.5\arcsec),
corresponding to an earlier phase when the mass-loss rate was about factor 3
higher than now. The near-IR image asymmetries discovered within the central
\about\ 0.1\arcsec of this system originate from \about\ \E{-4} \Mo\ of
swept-up wind material in a cocoon elongated along the axis, extending to a
radial distance of \about\ 1,100 AU. The cocoon expansion is driven by bipolar
jets that it confines, and that were switched on $\la$ 200 years ago. The axial
symmetry of the J-band image eliminates the possibility of a companion star,
unless closer than \about\ 5 stellar radii. Higher sensitivity and/or better
angular resolution would uncover image asymmetry in the K-band too.

Jet-driven cocoon expansion at various stages of development has
now been observed in a number of AGB stars, culminating in
breakout from the confining spherical wind (\S\ref{sec:sequence}).
The immediate post-AGB stage is believed to be the
proto-planetary-nebula (PPN) phase.  A morphological study of a
large sample of PPNs suggested the presence of jets that broke
through the massive AGB wind, but it was not known if a similar
morphology extends back to the AGB phase (Meixner et al.\ 1999;
Ueta, Meixner \& Bobrowsky 2000). Indeed, jets are found to be
quite common in PPN as shown by the recent observations of
IRAS16342-3814 (Sahai et al.\ 1999) K3-35 (Miranda et al.\ 2001),
Hen 3-1475 (Riera et al.\ 2003) and IRAS22036+5306 (Sahai et al
2003b), for example. The case of K3-35 is particularly striking
because of its great similarity to the AGB star W43A: water masers
at the tips of bipolar jets at a large distance from the systemic
center, which is surrounded by masers in the standard spherical
shell configuration. This strongly suggests that W43A provides a
glimpse of the immediate precursor of K3-35.

These new developments enable us to identify the first instance of symmetry
breaking in the evolution from AGB to planetary nebula. Bipolar asymmetry
appears during the final stages of AGB mass outflow. Mounting evidence suggests
that this asymmetry is driven by collimated outflow in the polar regions. More
complex geometries emerge in the post-AGB phase from a mixture of various
processes that could involve multiple jets, fast winds, etc. These processes
operate in the environment shaped by the AGB phase, leading to the myriad of
complex structures found in PPN sources (e.g. Su, Hrivnak \& Kwok 2001).

\section*{Acknowledgments}

We thank R. Sahai for most useful comments. Support by NSF grants PHY-0070928
(D.V.) and AST-0206149 (M.E.) is gratefully acknowledged. This work was also
supported by National Computational Science Alliance under AST020014 and
utilized the HP Superdome cluster at the University of Kentucky. We thank the
University of Kentucky's KAOS group at the Electrical Engineering department
for time on their 65-processor Linux cluster KLAT2.

\let\Ref=\item
{}

\appendix
\section {Details of Radiative Transfer Modeling}
\label{appendix}

\begin{figure}
\begin{center}
\framebox[3\width]{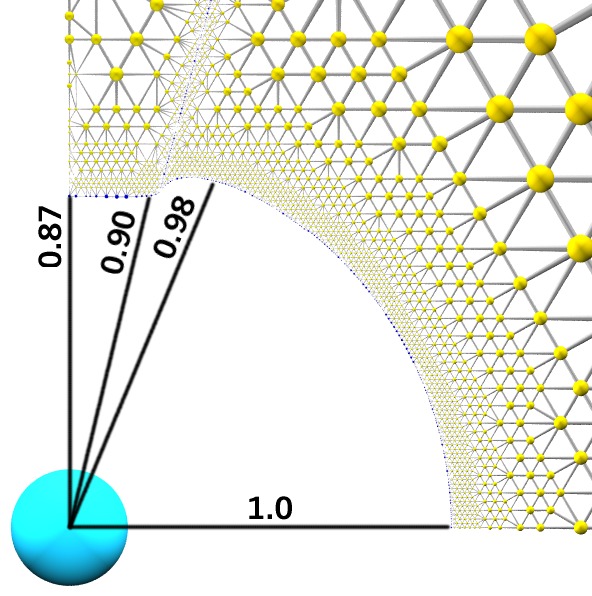}
\end{center}
\caption{\label{grid} A zoom into the central region of the
computational grid. Some radial dimensions of the dust-free cavity
are listed in terms of the dust condensation distance in the
equatorial plane. The stellar radius is $\rho_\star$ = 0.153.
Temperature is calculated at the grid points marked as spheres
(their sizes carry no particular meaning).}
\end{figure}

LELUYA \hbox{({\it www.leluya.org})} is our newly developed 2D radiative
transfer code that works with axially symmetric dust configurations. It solves
the integral equation of the formal solution of radiative transfer including
dust scattering, absorption and thermal emission. The solution is based on a
long-characteristics approach to the direct method of solving the matrix
version of the integral equation (Kurucz 1969). The equations are solved on a
highly unstructured triangular self-adaptive grid that enables LELUYA to
resolve simultaneously many orders of magnitude in both spatial and optical
depth space. It also enables automatic reshaping of the dust-free cavity around
the central source according to asymmetries in the diffuse radiation (eq. 1).
All grid points are coupled with each other through a correlation matrix based
on the dust scattering. A simple matrix inversion determines the solution of
radiative transfer for a given dust temperature distribution without any
iterations. The temperature is then updated and the procedure repeated.
Luminosity conservation within 5\% is achieved in only three steps.

\begin{figure}
\epsfig{figure=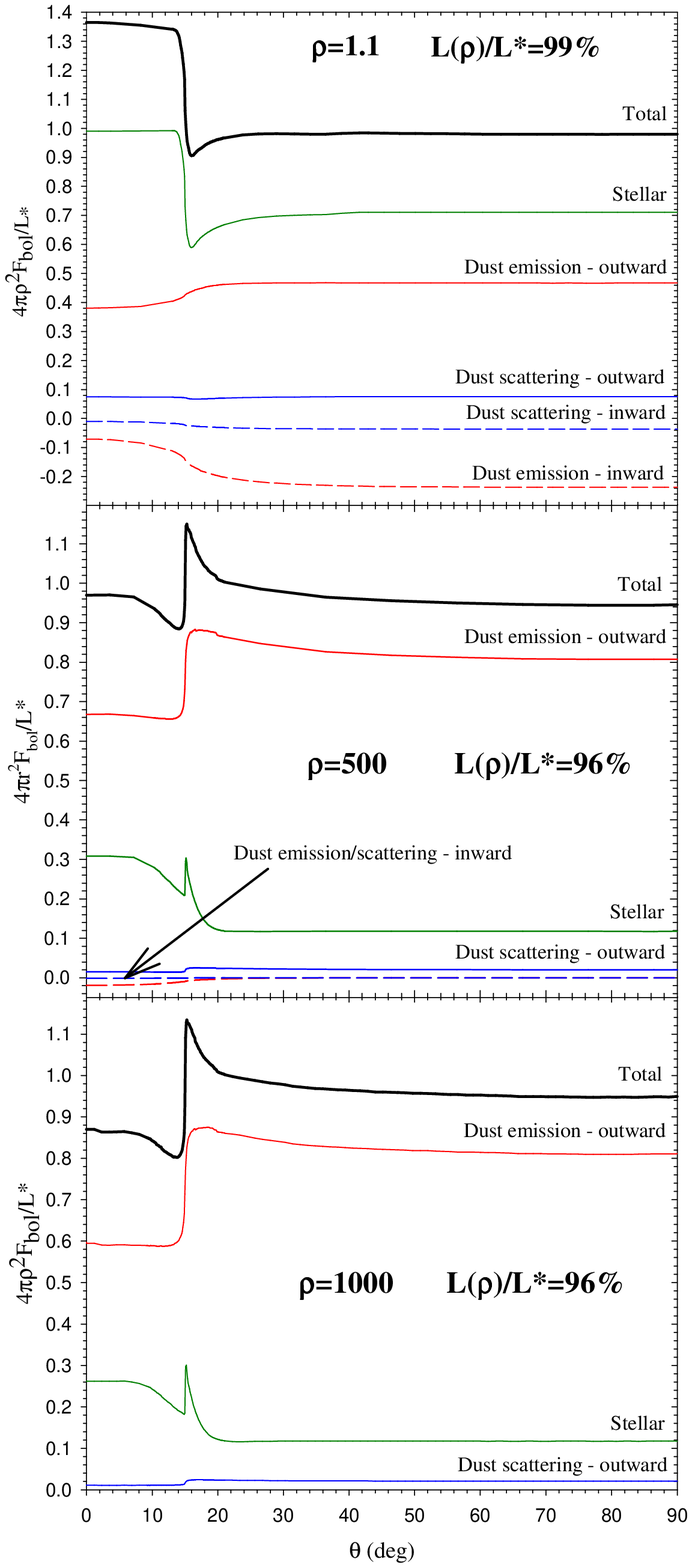,clip,width=\hsize}
\caption{\label{Fbol} Angular dependence of the radial bolometric
flux over spheres of radius $\rho$ = 1.1, 500 and 1000 for the
best fit minimal model described in \S\ref{sec:minimal}. The
numerical precision of luminosity conservation (eq.\ \ref{eq:L})
is indicated from the listed $L(\rho)/\L$ in each panel.}
\end{figure}

Figure \ref{grid} shows LELUYA's computational grid for the best fit minimal
model described in \S\ref{sec:minimal}. Three grids of different resolutions
were created for three sets of wavelengths, based on the density and optical
depth variation. The first grid has 2982 points and starts with $\tau^{\rm e} =
120$ at 0.2\mic, the shortest wavelength considered; this is the grid shown in
the figure. The second grid has 2836 points and starts at wavelengths with
$\tau^{\rm e}=1.2$. The third has 2177 grid points for wavelengths with
$\tau^{\rm e} \le 0.1$. Angular integration around a grid point is performed
over a highly non-uniform self-adaptive angular grid (with about 550 rays on
average).

\subsection{Luminosity Conservation}

Luminosity conservation is the test determining convergence to the correct
physical solution. A decrease in computed luminosity indicates energy sink due
to insufficient spatial grid resolution, while an increase reflects energy
excess due to a coarse angular grid. It is important to note that because of
the lack of spherical symmetry, {\em the bolometric flux does vary over
spherical surfaces}. The conserved quantity is luminosity, the energy
transmitted per unit time across any surface enclosing the star. For a sphere
of radius $\rho$, the luminosity is computed from the radial component of the
bolometric flux vector $F\sub{bol,r}$ via
\eq{\label{eq:L1}
   L(\rho) = 4\pi\rho^2\int_0^1 F\sub{bol,r}(\rho,\theta)\,d\cos\theta
}
and the luminosity conservation relation is
\eq{\label{eq:L}
 {L(\rho)\over \L} = 1
}
at every $\rho$, where \L\ is the stellar luminosity. In spherical symmetry
$F\sub{bol,r}$ is $\theta$-independent and $4\pi\rho^2F\sub{bol,r}/\L\ = 1$.
When the spherical symmetry is broken $F\sub{bol,r}$ becomes $\theta$-{\em
dependent} and $4\pi\rho^2F\sub{bol,r}(\theta)/\L$ can {\em exceed unity} in
certain directions, corresponding to locally enhanced energy outflow.

Our model calculations conserve luminosity within 5\% at all radii. Figure
\ref{Fbol} shows the angular variation of $F\sub{bol,r}(\theta)$ and its
following five contributions: stellar, inward and outward emission, and inward
and outward scattered flux. These angular variations are shown at $\rho$ = 1.1,
500 and 1000. The small spikes in $F\sub{bol,r}$ close to $\theta\sub{cone}$
are real, reflecting the irregular shape of the dust condensation surface. Even
though these irregularities are spatially small, their effect on optical depth
variations magnifies their importance. At small radii, energy outflow through
the cones is enhanced in comparison with the wind and is the main reason for
their higher temperature. This region is dominated by stellar contribution. At
large radii these roles are reversed, the diffuse radiation (mostly dust
emission) takes over and the temperatures inside and outside the cones become
equal.

\label{lastpage}
\end{document}